\documentclass[aps]{revtex4-1}%
\usepackage{amssymb}
\usepackage{amsfonts}
\usepackage{amsmath}
\usepackage{graphicx}%
\usepackage[T1]{fontenc}
\usepackage[utf8]{inputenc}
\begin{document}
\preprint{HEP/123-qed}
\title[Short title for running header]{Multipole matrix elements of Green function of Laplace equation}
\author{Karol Makuch}
\affiliation{Institute of Theoretical Physics, Faculty of Physics, University of Warsaw,
Pasteura 5, 02-093 Warsaw, Poland}
\email{Karol.Makuch@fuw.edu.pl}
\author{Przemys\l aw G\'{o}rka}
\affiliation{Department of Mathematics and Information Sciences, Warsaw University of
Technology, Koszykowa 75, 00-662 Warsaw, Poland}
\pacs{PACS number}

\begin{abstract}
Multipole matrix elements of Green function of Laplace equation are
calculated. The multipole matrix elements of Green function in electrostatics
describe potential on a sphere which is produced by a charge distributed on
the surface of a different (possibly overlapping) sphere of the same radius.
The matrix elements are defined by double convolution of two spherical
harmonics with the Green function of Laplace equation.
The method we use relies on the fact that in the Fourier space the double
convolution has simple form. Therefore we calculate the multipole matrix from
its Fourier transform. An important part of our considerations is
simplification of the three dimensional Fourier transformation of general
multipole matrix by its rotational symmetry to the one-dimensional Hankel transformation.

\end{abstract}
\volumeyear{year}
\volumenumber{number}
\issuenumber{number}
\eid{identifier}
\date[Date text]{date}
\received[Received text]{date}

\revised[Revised text]{date}

\accepted[Accepted text]{date}

\published[Published text]{date}

\startpage{1}
\endpage{102}
\maketitle

\section{Introduction}

Metallic spheres in a vacuum, minute spherical particles in a fluid, and
spherical inclusions in a solid body are often considered in statistical
physics of dispersive media \cite{sahimi2003heterogeneousI}. Equations which
describe those systems are linear and have spherical symmetry. It is the case
of Laplace equation in dielectrics \cite{griffiths1999introduction}, Stokes
equations in suspensions \cite{kim1991microhydrodynamics} and Lam\'{e}
equations \cite{sadd2014elasticity} describing solid body. The first step in
considerations of dispersive media is often to solve a single particle
problem. In dielectrics it means to find distribution of charge on the surface
of a single metallic sphere in an external electrostatic potential. To find
the solution of single particle problem it is convenient to take spherical
symmetry into account \cite{hinsen1992dielectric}. To this end one introduces
a set of scalar functions on the sphere which is invariant under rotations.
That is how spherical harmonics $Y_{lm}\left(  \mathbf{\hat{r}}\right)  $
enter calculations.

To pass from considerations of a single particle to the analysis of many
particles it demands to answer the following question. What is the
electrostatic potential produced by a charge distributed in one spherical
surface in the area occupied by a different sphere. The answer to the above
question can be inferred from the following multipole matrix element%
\begin{align}
&  \left[  G\left(  \mathbf{R},\mathbf{R}^{\prime}\right)  \right]
_{lm,l^{\prime}m^{\prime}}\nonumber\\
&  =\int_{\mathcal{R}^{3}}d^{3}r\int_{\mathcal{R}^{3}}d^{3}r^{\prime}\frac
{1}{a}\delta\left(  \left\vert \mathbf{r}-\mathbf{R}\right\vert -a\right)
\phi_{lm}^{\ast}\left(  \mathbf{r}-\mathbf{R}\right)  \mathbf{G}\left(
\mathbf{r}-\mathbf{r}^{\prime}\right)  \frac{1}{a}\delta\left(  \left\vert
\mathbf{r}^{\prime}-\mathbf{R}^{\prime}\right\vert -a\right)  \phi_{l^{\prime
}m^{\prime}}\left(  \mathbf{r}^{\prime}-\mathbf{R}^{\prime}\right)
.\label{def G multi}%
\end{align}
In the above definition $\mathbf{G}\left(  \mathbf{r}\right)  $ is the Green
function of Laplace equation \cite{griffiths1999introduction},%
\begin{equation}
\mathbf{G}\left(  \mathbf{r}\right)  =\frac{1}{4\pi r}%
,\label{Laplace eq Green function} \nonumber%
\end{equation}
$a$ is radius of particles, and one-dimensional Dirac delta distribution
$\delta\left(  x\right)  $ is used. Moreover $\phi_{lm}\left(  \mathbf{r}%
\right)  $ are solid harmonics defined by%
\begin{equation}
\phi_{lm}\left(  \mathbf{r}\right)  =r^{l}Y_{lm}\left(  \mathbf{\hat{r}%
}\right)  , \nonumber
\end{equation}
with spherical harmonics $Y_{lm}\left(  \mathbf{\hat{r}}\right)  $ numbered by
order $l=0,1,\ldots$ and azimuthal number $m=-l,\ldots,l$
\cite{edmonds1996angular}. In our notation an argument of spherical harmonics
is a versor $\mathbf{\hat{r}}\left(  \theta,\phi\right)  $ in direction
described by angles $\theta,\phi$ in spherical coordinates. Dirac delta
distributions in equation (\ref{def G multi}) reduce three-dimensional
integrations to integrals over the surface of the spheres with radius $a$
centered at positions $\mathbf{R}$ and $\mathbf{R}^{\prime}$.

We can differentiate between two situations. The first situation corresponds
to nonoverlapping spheres, i.e. when $\left\vert \mathbf{R}-\mathbf{R}%
^{\prime}\right\vert >2a$. In this case the matrix elements defined by
equation (\ref{def G multi}) can be inferred from the results in the
literature \cite{sheu1990simulation}. They have application e.g. in numerical
simulations. The second situation is the case of overlapping configurations,
$\left\vert \mathbf{R}-\mathbf{R}^{\prime}\right\vert <2a$. Even if the
particles in the system cannot overlap there may appear a need of calculation
of overlapping configurations of the multipole matrix elements $\left[
G\left(  \mathbf{R},\mathbf{R}^{\prime}\right)  \right]  _{lm,l^{\prime
}m^{\prime}}$. For example, it has been recognized that overlapping
configurations appear in microscopic explanation of the famous
Clauius-Mossotti formula for dielectric constant \cite{felderhof1983clausius}.
In this context integral $\int_{\left\vert \mathbf{R}-\mathbf{R}^{\prime
}\right\vert <2a}d^{3}R^{\prime}\left[  G\left(  \mathbf{R},\mathbf{R}%
^{\prime}\right)  \right]  \exp\left(  -i\mathbf{kR}^{\prime}\right)
_{lm,l^{\prime}m^{\prime}}$ for low multipole numbers $l,l^{\prime}$ has been
considered. Therefore the multipole matrix elements $\left[  G\left(
\mathbf{R},\mathbf{R}^{\prime}\right)  \right]  _{lm,l^{\prime}m^{\prime}}$
for overlapping configurations play an important role in statistical physics
considerations of dispersive media.

In this article we give general expression for the multipole matrix elements
$\left[  G\left(  \mathbf{R},\mathbf{R}^{\prime}\right)  \right]
_{lm,l^{\prime}m^{\prime}}$ defined by equation (\ref{def G multi}). The new
contribution of the current article is the calculation of $\left[  G\left(
\mathbf{R},\mathbf{R}^{\prime}\right)  \right]  _{lm,l^{\prime}m^{\prime}}$
for overlapping configurations $\left\vert \mathbf{R}-\mathbf{R}^{\prime
}\right\vert <2a$. We are going to use the result for overlapping
configurations in our statistical physics considerations of transport
properties of dispersive media in further work.

The method of calculation of the multipole matrix elements defined by equation
(\ref{def G multi}) is the following. We observe that $\left[  G\left(
\mathbf{R},\mathbf{R}^{\prime}\right)  \right]  _{lm,l^{\prime}m^{\prime}}$
has the form of a double convolution of three functions - two solid harmonics
and the Green function. Therefore we calculate the Fourier transform of the
three functions, take their product, and then perform the inverse Fourier
transform to obtain $G\left(  \mathbf{R},\mathbf{R}^{\prime}\right)  $. An
important element of our calculations is to use spherical symmetry which
allows to reduce the three-dimensional Fourier transform of the multipole
matrix to the one-dimensional Hankel transform.

\section{Multipole matrix in the Fourier space}

The aim of this article is calculation of integral (\ref{def G multi}).
We can simplify it using homogeneity of the Laplace equation which implies
that the multipole matrix $\left[  G\left(  \mathbf{R},\mathbf{R}^{\prime
}\right)  \right]  _{lm,l^{\prime}m^{\prime}}$ depends on the relative
positions $\mathbf{R-R}^{\prime}$. Therefore from now on we consider multipole
matrix $\left[  G\left(  \mathbf{R}\right)  \right]  _{lm,l^{\prime}m^{\prime
}}$ defined by%

\begin{equation}
\left[  G\left(  \mathbf{R}-\mathbf{R}^{\prime}\right)  \right]
_{lm,l^{\prime}m^{\prime}}=\left[  G\left(  \mathbf{R},\mathbf{R}^{\prime
}\right)  \right]  _{lm,l^{\prime}m^{\prime}} \nonumber
\end{equation}
depending on the relative positions.

Next, we observe that the multipole matrix given by the formula
(\ref{def G multi}) has a form of a double convolution of three functions. The
three functions are $\omega_{l^{\prime}m^{\prime}}\left(  \mathbf{r}\right)
\equiv\delta\left(  \left\vert \mathbf{r}\right\vert -a\right)  \phi
_{l^{\prime}m^{\prime}}\left(  \mathbf{r}\right)  /a$, conjugated function
$\omega_{lm}^{\ast}\left(  \mathbf{r}\right)  $, and the Green function
$\mathbf{G}\left(  \mathbf{r}\right)  $. The Fourier transform of a double
convolution of the three functions is given by the product of their Fourier
transforms, therefore%
\begin{equation}
\left[  \hat{G}\left(  \mathbf{k}\right)  \right]  _{lm,l^{\prime}m^{\prime}%
}=\hat{\omega}_{lm}^{\ast}\left(  \mathbf{k}\right)  \mathbf{\hat{G}}\left(
\mathbf{k}\right)  \hat{\omega}_{l^{\prime}m^{\prime}}\left(  \mathbf{k}%
\right). \nonumber
\end{equation}
In our calculations we use the following definition of the three-dimensional
Fourier transform%

\begin{equation}
\hat{G}_{lm,l^{\prime}m^{\prime}}\left(  \mathbf{k}\right)  =\int
d^{3}R\mathbf{\ }\exp\left(  -i\mathbf{kR}\right)  G_{lm,l^{\prime}m^{\prime}%
}\left(  \mathbf{R}\right)  , \label{def Fourier}%
\end{equation}
with the inverse transformation given by the formula%

\begin{equation}
G_{lm,l^{\prime}m^{\prime}}\left(  \mathbf{R}\right)  =\frac{1}{\left(
2\pi\right)  ^{3}}\int d^{3}k\mathbf{\ }\exp\left(  i\mathbf{kR}\right)
\hat{G}_{lm,l^{\prime}m^{\prime}}\left(  \mathbf{k}\right)  .
\label{def inv Fourier} \nonumber%
\end{equation}
In this situation the Fourier transforms of $\omega_{lm}\left(  \mathbf{r}%
\right)  $ and $G\left(  \mathbf{r}\right)  $ are given respectively by%

\begin{equation}
\mathbf{\hat{G}}\left(  \mathbf{k}\right)  =\frac{1}{k^{2}}, \nonumber
\end{equation}
and%
\begin{equation}
\hat{\omega}_{lm}\left(  \mathbf{k}\right)  =4\pi a^{l+1}\left(  -i\right)
^{l}j_{l}\left(  ka\right)  Y_{lm}\left(  \mathbf{\hat{k}}\right)  . \nonumber
\end{equation}
The last expression can be calculated with the use of equation (5.8.3) from
Ref. \cite{edmonds1996angular} and with the use of orthonormality of spherical
harmonics. The expression contains spherical Bessel functions $j_{l}\left(
x\right)  $ of the order $l$. Finally, the Fourier transform of multipole
matrix (\ref{def G multi}) is given by the following formula%

\begin{equation}
\left[  \hat{G}\left(  \mathbf{k}\right)  \right]  _{lm,l^{\prime}m^{\prime}%
}=\left(  4\pi\right)  ^{2}\left(  -i\right)  ^{-l+l^{\prime}}a^{l+l^{\prime
}+2}\frac{j_{l}\left(  ka\right)  j_{l^{\prime}}\left(  ka\right)  }{k^{2}%
}Y_{lm}^{\ast}\left(  \mathbf{\hat{k}}\right)  Y_{l^{\prime}m^{\prime}}\left(
\mathbf{\hat{k}}\right)  . \label{G fourier transform}%
\end{equation}
According to the above expression, the Fourier transform of $\left[  G\left(
\mathbf{R}\right)  \right]  _{lm,l^{\prime}m^{\prime}}$ is given by the
spherical harmonics and the spherical Bessel functions. At this point we face
the main difficulty in our calculations of $\left[  G\left(  \mathbf{R}%
\right)  \right]  _{lm,l^{\prime}m^{\prime}}$. The inverse Fourier transform
of the above formula has to be calculated. To this end we consider rotational
symmetry of the multipole matrix.

\section{Rotational symmetry of a multipole matrix}

In the definition of the multipole matrix (\ref{def G multi}) there appear
solid harmonics $\phi_{lm}\left(  \mathbf{r}\right)  $ and Green function
$\mathbf{G}\left(  \mathbf{r}\right)  $. Transformation of solid harmonics
under rotation is given by the formula%
\begin{equation}
\phi_{lm}\left(  \mathbf{D}\left(  \alpha,\gamma,\beta\right)  \mathbf{r}%
\right)  =\sum_{m_{1}=-l}^{l}\left[  D^{\left(  l\right)  }\left(
\alpha,\gamma,\beta\right)  \right]  _{mm_{1}}^{\ast}\phi_{lm_{1}}\left(
\mathbf{r}\right)  .
\end{equation}
The above expression can be inferred from Ref. \cite{edmonds1996angular} from
which we adopt notation in this article. There, formula (4.1.1) defines
three-dimensional rotation matrix $\mathbf{D}\left(  \alpha,\beta
,\gamma\right)  $ characterized by the Euler angles $\alpha,\beta,\gamma$.
Moreover, $D_{mm^{\prime}}^{\left(  l\right)  }\left(  \alpha,\beta
,\gamma\right)  $ denotes the Wigner matrix. Isotropy of the Laplace equation
implies that its Green function $\mathbf{G}\left(  \mathbf{r}\right)  $ is
invariant under rotation, i.e. $\mathbf{G}\left(  \mathbf{D}\left(
\alpha,\beta,\gamma\right)  \mathbf{r}\right)  =\mathbf{G}\left(
\mathbf{r}\right)  $.

Changing the variables in the integrals in Eq. (\ref{def G multi}) and the
above properties of the solid harmonics and the Green function lead us to the
following transformation of the multipole matrix elements under rotation%
\begin{equation}
G_{lm,l^{\prime}m^{\prime}}\left(  \mathbf{D}\left(  \alpha,\beta
,\gamma\right)  \mathbf{R}\right)  =\sum_{m_{1},m_{1}^{\prime}}D_{m,m_{1}%
}^{\left(  l\right)  }\left(  \alpha,\beta,\gamma\right)  \left[
D_{m^{\prime},m_{1}^{\prime}}^{\left(  l^{\prime}\right)  }\left(
\alpha,\beta,\gamma\right)  \right]  ^{\ast}G_{lm_{1},l^{\prime}m_{1}^{\prime
}}\left(  \mathbf{R}\right)  .\label{G rot sym}%
\end{equation}
It is worth mentioning here that the above transformation applied for
$\mathbf{R}$ in $z$ direction, $\mathbf{R}=R\mathbf{\hat{e}}_{z}$, and for any
rotations around axis $z$, thus for $\beta=0$ and any other Euler angles,
implies that the multipole matrix is diagonal in indexes $m$, i.e.%
\begin{equation}
G_{lm,l^{\prime}m^{\prime}}\left(  R\mathbf{\hat{e}}_{z}\right)
=\delta_{m,m^{\prime}}G_{lm,l^{\prime}m}\left(  R\mathbf{\hat{e}}_{z}\right)
, \nonumber
\end{equation}
where Kronecker delta $\delta_{m,m^{\prime}}$ appears. It is  easy to prove
similar diagonality in case of the Fourier transform,%
\begin{equation}
\hat{G}_{lm,l^{\prime}m^{\prime}}\left(  k\mathbf{\hat{e}}_{z}\right)
=\delta_{m,m^{\prime}}\hat{G}_{lm,l^{\prime}m}\left(  k\mathbf{\hat{e}}%
_{z}\right)  ,\label{fourier m diagonality}%
\end{equation}
which appears as a result of Fourier transformation of equation
(\ref{G rot sym}) and considerations for proper rotations and wave vector
$\mathbf{k}=k\mathbf{\hat{e}}_{z}$.

\section{Fourier transform of multipole matrix - simplification by rotational
symmetry}

The key point of our calculations is simplification of the Fourier transform
of a matrix satisfying symmetry property given by equation (\ref{G rot sym}).
We simplify the Fourier transform for a wave vector along $z$ direction
because then the multipole matrix is diagonal in indexes $m$, as is shown by
equation (\ref{fourier m diagonality}). For the case $\mathbf{k}%
=k\mathbf{\hat{e}}_{z}$ expression (\ref{def Fourier}) written in spherical
coordinates reads%
\begin{equation}
\hat{G}_{lm,l^{\prime}m^{\prime}}\left(  k\mathbf{\hat{e}}_{z}\right)
=\int_{0}^{\infty}dR\int_{0}^{\pi}d\theta\int_{0}^{2\pi}d\phi R^{2}\sin
\theta\exp\left(  -ikR\cos\theta\right)  G_{lm,l^{\prime}m^{\prime}}\left(
\mathbf{R}\left(  R,\theta,\phi\right)  \right)  .\label{pom trans 1}%
\end{equation}
We express vector $\mathbf{R}\left(  R,\theta,\phi\right)  $ by a product of
the vector $R\mathbf{\hat{e}}_{z}$ and the rotation matrix $\mathbf{D}\left(
\alpha,\beta,\gamma\right)  $ characterized with proper Euler angles. The
angles can be deduced from formula (4.1.1) in Ref. \cite{edmonds1996angular}.
These angles are $\beta=-\theta$, $\gamma=-\phi$, and any $\alpha$, e.g.
$\alpha=0$. Therefore $\mathbf{R}\left(  R,\theta,\phi\right)  =\mathbf{D}%
\left(  0,-\theta,-\phi\right)  R\mathbf{\hat{e}}_{z}$. For this rotation and
$\mathbf{R=}R\mathbf{\hat{e}}_{z}$ we use symmetry property (\ref{G rot sym})
which leads to the following expression for multipole matrix $G$%
\begin{equation}
G_{lm,l^{\prime}m^{\prime}}\left(  \mathbf{R}\left(  R,\theta,\phi\right)
\right)  =\sum_{m_{1},m_{1}^{\prime}}D_{m,m_{1}}^{\left(  l\right)  }\left(
0,-\theta,-\phi\right)  \left[  D_{m^{\prime},m_{1}^{\prime}}^{\left(
l^{\prime}\right)  }\left(  0,-\theta,-\phi\right)  \right]  ^{\ast}%
G_{lm_{1},l^{\prime}m_{1}^{\prime}}\left(  R\mathbf{\hat{e}}_{z}\right)
.\label{G for any R by R ez}%
\end{equation}

In the next step we represent $\exp\left(  -ikR\cos\theta\right)  $ from
expression (\ref{pom trans 1}) in the form of an infinite series of spherical
Bessel functions%
\begin{equation}
\exp\left(  -ikR\cos\theta\right)  =\sum_{l_{1}=0}^{\infty}\left(  -i\right)
^{l_{1}}\left(  2l_{1}+1\right)  j_{l_{1}}\left(  kR\right)  D_{00}^{\left(
l_{1}\right)  }\left(  0,-\theta,-\phi\right)  ,
\label{plane wave toward z direction} \nonumber%
number
\end{equation}
which is deduced from formula (5.8.1) and (4.1.26) in the reference
\cite{edmonds1996angular}. Taking into consideration the last two formulae in
expression (\ref{pom trans 1}) yields%

\begin{align}
\hat{G}_{lm,l^{\prime}m^{\prime}}\left(  k\mathbf{\hat{e}}_{z}\right)   &
=\int_{0}^{\infty}dRR^{2}\sum_{l_{1}=0}^{\infty}\left(  -i\right)  ^{l_{1}%
}\left(  2l_{1}+1\right)  j_{l_{1}}\left(  kR\right)  G_{lm_{1},l^{\prime
}m_{1}^{\prime}}\left(  R\mathbf{\hat{e}}_{z}\right)  \times\nonumber\\
&  \sum_{m_{1},m_{1}^{\prime}}\int_{0}^{\pi}d\theta\int_{0}^{2\pi}d\phi
\sin\theta D_{00}^{\left(  l_{1}\right)  }\left(  0,\theta,\phi\right)
D_{m,m_{1}}^{\left(  l\right)  }\left(  0,\theta,\phi\right)  \left[
D_{m^{\prime},m_{1}^{\prime}}^{\left(  l^{\prime}\right)  }\left(
0,\theta,\phi\right)  \right]  ^{\ast}.\label{pom 2}%
\end{align}
Integration over variables $\theta,\phi$ in the above formula is performed
with the use of formulae (4.6.2), (4.1.12) and (4.2.7) from the reference
\cite{edmonds1996angular}. They lead to expression%
\begin{align}
&  \int_{0}^{\pi}d\theta\int_{0}^{2\pi}d\phi\sin\theta D_{00}^{\left(
l_{1}\right)  }\left(  0,\theta,\phi\right)  D_{m,m_{1}}^{\left(  l\right)
}\left(  0,\theta,\phi\right)  \left[  D_{m^{\prime},m_{1}^{\prime}}^{\left(
l^{\prime}\right)  }\left(  0,\theta,\phi\right)  \right]  ^{\ast} \nonumber \\
&  =4\pi\left(  -1\right)  ^{m^{\prime}-m_{1}^{\prime}}\left(
\begin{array}
[c]{ccc}%
l & l^{\prime} & l_{1}\\
m & -m^{\prime} & 0
\end{array}
\right)  \left(
\begin{array}
[c]{ccc}%
l & l^{\prime} & l_{1}\\
m_{1} & -m_{1}^{\prime} & 0
\end{array}
\right)  ,\nonumber
\end{align}
which contains Wigner $3$-$j$ symbols. Taking into account the above integral
in expression (\ref{pom 2}) yields%
\begin{align}
\hat{G}_{lm,l^{\prime}m^{\prime}}\left(  k\mathbf{\hat{e}}_{z}\right)   &
=4\pi\sum_{l_{1}=\left\vert l-l^{\prime}\right\vert }^{\left\vert l+l^{\prime
}\right\vert }\sum_{m_{1}=-l}^{l}\sum_{m_{1}^{\prime}=-l^{\prime}}^{l^{\prime
}}\left(  -1\right)  ^{m^{\prime}-m_{1}^{\prime}}\left(  2l_{1}+1\right)
\left(
\begin{array}
[c]{ccc}%
l & l^{\prime} & l_{1}\\
m & -m^{\prime} & 0
\end{array}
\right)  \left(
\begin{array}
[c]{ccc}%
l & l^{\prime} & l_{1}\\
m_{1} & -m_{1}^{\prime} & 0
\end{array}
\right)  \times\nonumber\\
&  \left(  -i\right)  ^{l_{1}}\int_{0}^{\infty}dRR^{2}j_{l_{1}}\left(
kR\right)  G_{lm_{1},l^{\prime}m_{1}^{\prime}}\left(  R\mathbf{\hat{e}}%
_{z}\right)  .\label{transf four zredukowana}%
\end{align}
In this way the three-dimensional Fourier transform is reduced to the one
dimensional Hankel transform \cite{bleistein1975asymptotic}.

It is convenient to introduce different representation of matrix $G$. Namely,
instead of $G_{lm,l^{\prime}m^{\prime}}\left(  R\mathbf{\hat{e}}_{z}\right)  $
we can consider $g_{l,l^{\prime}}^{j}\left(  R\right)  $ defined in the
following way%
\begin{equation}
g_{l,l^{\prime}}^{j}\left(  R\right)  =\left(  2j+1\right)  \sum_{m,m^{\prime
}}\left(  -1\right)  ^{m}\left(
\begin{array}
[c]{ccc}%
l & l^{\prime} & j\\
m & -m^{\prime} & 0
\end{array}
\right)  G_{lm,l^{\prime}m^{\prime}}\left(  R\mathbf{\hat{e}}_{z}\right)  .
\label{from canonical basis to J}%
\end{equation}
with the inverse transformation%
\begin{equation}
G_{lm,l^{\prime}m^{\prime}}\left(  R\mathbf{\hat{e}}_{z}\right)
=\delta_{m,m^{\prime}}\left(  -1\right)  ^{m}\sum_{j=\left\vert l-l^{\prime
}\right\vert }^{l+l^{\prime}}\left(
\begin{array}
[c]{ccc}%
l & l^{\prime} & j\\
m & -m^{\prime} & 0
\end{array}
\right)  g_{l,l^{\prime}}^{j}\left(  R\right)  .
\label{from J basis to canonical}%
\end{equation}
Let us notice that only integer $j$ which satisfy condition $\left\vert
l-l^{\prime}\right\vert \leq j\leq l+l^{\prime}$ needs to be considered in the
above equations. It follows from properties of Wigner $3$-$j$ symbols. The
multipole matrix elements $G_{lm,l^{\prime}m^{\prime}}\left(  \mathbf{R}%
\right)  $ for any $\mathbf{R}$ are then related to $g_{l,l^{\prime}}%
^{j}\left(  R\right)  $ by equation%
\begin{equation}
G_{lm,l^{\prime}m^{\prime}}\left(  \mathbf{R}\left(  R,\theta,\phi\right)
\right)  =\sum_{m_{1}}\left(  -1\right)  ^{m^{\prime}+m_{1}}\sum_{j=\left\vert
l-l^{\prime}\right\vert }^{l+l^{\prime}}\left(
\begin{array}
[c]{ccc}%
l & l^{\prime} & j\\
m & -m^{\prime} & m_{1}%
\end{array}
\right)  D_{-m_{1},0}^{\left(  j\right)  }\left(  0,-\theta,-\phi\right)
g_{l,l^{\prime}}^{j}\left(  R\right)  ,
\end{equation}
which follows from equation (\ref{G for any R by R ez}) and relation
(\ref{from J basis to canonical}).

Different representations of the multipole matrix $G_{lm,l^{\prime}m^{\prime}%
}\left(  R\mathbf{\hat{e}}_{z}\right)  $ in positional space given by
equations (\ref{from canonical basis to J}) and
(\ref{from J basis to canonical}) can be similarly introduced also in Fourier
space, i.e.%

\begin{equation}
\tilde{g}_{l,l^{\prime}}^{j}\left(  k\right)  =\left(  2j+1\right)
\sum_{m,m^{\prime}}\left(  -1\right)  ^{m}\left(
\begin{array}
[c]{ccc}%
l & l^{\prime} & j\\
m & -m^{\prime} & 0
\end{array}
\right)  \hat{G}_{lm,l^{\prime}m^{\prime}}\left(  k\mathbf{\hat{e}}%
_{z}\right)  .\label{from canonical basis to J fourier}%
\end{equation}%
\begin{equation}
\hat{G}_{lm,l^{\prime}m^{\prime}}\left(  k\mathbf{\hat{e}}_{z}\right)
=\delta_{m,m^{\prime}}\left(  -1\right)  ^{m}\sum_{j=\left\vert l-l^{\prime
}\right\vert }^{l+l^{\prime}}\left(
\begin{array}
[c]{ccc}%
l & l^{\prime} & j\\
m & -m^{\prime} & 0
\end{array}
\right)  \tilde{g}_{l,l^{\prime}}^{j}\left(  k\right)
.\label{from J basis to canonical Fourier}%
\end{equation}

By equations (\ref{transf four zredukowana}),
(\ref{from canonical basis to J fourier}), and
(\ref{from J basis to canonical}) and orthogonality of Wigner $3$-$j$ symbols
\cite{edmonds1996angular}, we get that in the new basis the Fourier transform
of the multipole matrix is expressed in the following form%
\begin{equation}
\tilde{g}_{l,l^{\prime}}^{j}\left(  k\right)  =4\pi\left(  -i\right)  ^{j}%
\int_{0}^{\infty}dRR^{2}j_{j}\left(  kR\right)  g_{l,l^{\prime}}^{j}\left(
R\right)  .\label{Fourier reduced}%
\end{equation}
In this way the three-dimensional Fourier transform given by expression
(\ref{def Fourier}) of a multipole matrix satisfying rotational symmetry
(\ref{G rot sym}) is reduced to relevant Hankel transform given by expression
(\ref{Fourier reduced}).

Calculations performed in this section can be repeated with very minor
modification in order to reduce the inverse three-dimensional Fourier
transform of a multipole matrix to the one-dimensional Hankel transform. We
omit the derivation giving only the result for the inverse Fourier transform
of a multipole matrix satisfying rotational symmetry,%
\begin{equation}
g_{l,l^{\prime}}^{j}\left(  R\right)  =\frac{1}{2\pi^{2}}i^{j}\int_{0}%
^{\infty}dk\ k^{2}j_{j}\left(  kR\right)  \tilde{g}_{l,l^{\prime}}^{j}\left(
k\right)  .\label{inv four transform J basisi}%
\end{equation}

\section{Multipole matrix in positional space}

To calculate the inverse Fourier transform of $\left[  \hat{G}\left(
\mathbf{k}\right)  \right]  _{lm,l^{\prime}m^{\prime}}$ with the use of
equation (\ref{inv four transform J basisi}), $\tilde{g}_{ll^{\prime}}%
^{j}\left(  k\right)  $ is needed. We calculate it by
means of transformation (\ref{from canonical basis to J fourier}) and
expression (\ref{G fourier transform}) for $\hat{G}_{lml^{\prime}m^{\prime}%
}\left(  k\mathbf{\hat{e}}_{z}\right)  $. The calculations yields%
\begin{equation}
\tilde{g}_{l,l^{\prime}}^{j}\left(  k\right)  =4\pi\left(  -i\right)
^{-l+l^{\prime}}\left(  2j+1\right)  \left[  \left(  2l+1\right)  \left(
2l^{\prime}+1\right)  \right]  ^{1/2}a^{l+l^{\prime}+2}\left(
\begin{array}
[c]{ccc}%
l & l^{\prime} & j\\
0 & 0 & 0
\end{array}
\right)  \frac{j_{l}\left(  ka\right)  j_{l^{\prime}}\left(  ka\right)
}{k^{2}}.
\end{equation}
It is worth noting that the Wigner $3-j$ symbol is not vanishing only when the
$l,l^{\prime},j$ satisfy triangular inequality, $\left\vert l-l^{\prime
}\right\vert \leq j\leq\left\vert l+l^{\prime}\right\vert $. In calculations
of the above formula we used the following property of the spherical
harmonics, $Y_{jm}\left(  \mathbf{\hat{e}}_{z}\right)  =\delta_{m,0}\left(
\left(  2j+1\right)  /4\pi\right)  ^{1/2}$.

With the above expression for $\tilde{g}_{l,l^{\prime}}^{j}\left(  k\right)
$, equation (\ref{inv four transform J basisi}) yields%
\begin{equation}
g_{l,l^{\prime}}^{j}\left(  R\right)  =\mu_{j,l,l^{\prime}}a^{l+l^{\prime}%
+2}\int_{0}^{\infty}dk\ j_{j}\left(  kR\right)  j_{l}\left(  ka\right)
j_{l^{\prime}}\left(  ka\right)  \label{Oseen mult in R pom}%
\end{equation}
with%
\begin{equation}
\mu_{j,l,l^{\prime}}=\frac{2\left(  -i\right)  ^{-l+l^{\prime}+j}\left(
-1\right)  ^{j}}{\pi}\left(  2j+1\right)  \left[  \left(  2l+1\right)  \left(
2l^{\prime}+1\right)  \right]  ^{1/2}\left(
\begin{array}
[c]{ccc}%
l & l^{\prime} & j\\
0 & 0 & 0
\end{array}
\right)  .\label{mu coef}%
\end{equation}
It demands to calculate integral of three spherical Bessel functions. That has
already been considered in the literature \cite{prudnikov1983integrals}. For
$R\geq2a$, the integral is given as follows%
\begin{align}
&  \int_{0}^{\infty}dk\ j_{l_{1}}\left(  kR\right)  j_{l}\left(  ka\right)
j_{l^{\prime}}\left(  ka\right)  \nonumber\\
&  =\frac{\pi^{3/2}}{8a}\delta_{l+l^{\prime},l_{1}}\left(  \frac{a}{R}\right)
^{l+l^{\prime}+1}\frac{\Gamma\left(  \frac{1}{2}+l+l^{\prime}\right)  }%
{\Gamma\left(  \frac{3}{2}+l\right)  \Gamma\left(  \frac{3}{2}+l^{\prime
}\right)  },\label{int three bessels nonov}%
\end{align}
with Euler Gamma function $\Gamma\left(  x\right)  $. Whereas for $R\leq2a$ we
have%
\begin{align}
&  \int_{0}^{\infty}dk\ j_{j}\left(  kR\right)  j_{l}\left(  ka\right)
j_{l^{\prime}}\left(  ka\right)  \nonumber\\
&  =\frac{\pi^{3/2}}{2a}\frac{R}{a}\alpha_{j,l,l^{\prime}}\ _{4}F_{3}\left(
\frac{-l-l^{\prime}}{2},\frac{1+l-l^{\prime}}{2},\frac{1-l+l^{\prime}}%
{2},\frac{2+l+l^{\prime}}{2};\frac{1}{2},\frac{3-j}{2},\frac{4+j}{2}%
;\frac{R^{2}}{4a^{2}}\right)  \nonumber\\
&  +\frac{\pi^{3/2}}{2a}\left(  \frac{R}{a}\right)  ^{j}\beta_{j,l,l^{\prime}%
}\ _{4}F_{3}\left(  \frac{j-l-l^{\prime}-1}{2},\frac{j+l-l^{\prime}}{2}%
,\frac{j+l^{\prime}-l}{2},\frac{l+l^{\prime}+j+1}{2};\frac{1+j}{2},\frac{j}%
{2},\frac{3}{2}+j;\frac{R^{2}}{4a^{2}}\right)  \nonumber\\
&  -\frac{\pi^{3/2}}{2a}\frac{R^{2}}{a^{2}}\gamma_{j,l,l^{\prime}}\ _{4}%
F_{3}\left(  1-\frac{l^{\prime}+l+1}{2},1+\frac{l-l^{\prime}}{2}%
,1+\frac{l^{\prime}-l}{2},1+\frac{l+l^{\prime}+1}{2};\frac{3}{2},2-\frac{j}%
{2},2+\frac{j+1}{2};\frac{R^{2}}{4a^{2}}\right)  \label{int three bessels ov}%
\end{align}
with coefficients $\alpha_{j,l,l^{\prime}}$, $\beta_{j,l,l^{\prime}}$, and
$\gamma_{j,l,l^{\prime}}$ given by%
\begin{align}
\alpha_{j,l,l^{\prime}} &  =2^{-5/2}\frac{\Gamma\left(  \frac{j-1}{2}\right)
}{\Gamma\left(  \frac{1+l^{\prime}-l}{2}\right)  \Gamma\left(  \frac
{1+l-l^{\prime}}{2}\right)  \Gamma\left(  \frac{j+4}{2}\right)  },\nonumber\\
\beta_{j,l,l^{\prime}} &  =2^{-3/2}\frac{\Gamma\left(  1-j\right)
\Gamma\left(  \frac{1+l+l^{\prime}+j}{2}\right)  }{\Gamma\left(
1+\frac{1+l+l^{\prime}-j}{2}\right)  \Gamma\left(  1+\frac{l^{\prime}-l-j}%
{2}\right)  \Gamma\left(  1+\frac{l-l^{\prime}-j}{2}\right)  \Gamma\left(
\frac{3}{2}+j\right)  },\nonumber\\
\gamma_{j,l,l^{\prime}} &  =-2^{-7/2}\frac{\Gamma\left(  \frac{j}{2}-1\right)
}{\Gamma\left(  \frac{l^{\prime}-l}{2}\right)  \Gamma\left(  \frac
{l-l^{\prime}}{2}\right)  \Gamma\left(  2+\frac{j+1}{2}\right)  }%
.\label{abc coef}%
\end{align}
Symbol $_{4}F_{3}$ stands for hypergeometric function%
\begin{equation}
_{4}F_{3}\left(  \alpha_{1},\alpha_{2},\alpha_{3},\alpha_{4};\beta_{1}%
,\beta_{2},\beta_{3};x\right)  =\sum_{k=0}^{\infty}\frac{\left(  \alpha
_{1}\right)  _{k}\left(  \alpha_{2}\right)  _{k}\left(  \alpha_{3}\right)
_{k}\left(  \alpha_{4}\right)  _{k}}{\left(  \beta_{1}\right)  _{k}\left(
\beta_{2}\right)  _{k}\left(  \beta_{3}\right)  _{k}}\frac{x^{k}}{k!} \nonumber
\end{equation}
where $\left(  \alpha\right)  _{k}$ denotes Pochhammer symbol defined by%
\begin{equation}%
\begin{array}
[c]{cc}%
\left(  \alpha\right)  _{0}=1 & \text{for }k=0\\
\left(  \alpha\right)  _{k}=\alpha\left(  \alpha+1\right)  \ldots\left(
\alpha+k-1\right)   & \text{for }k=1,2,\ldots
\end{array}
. \nonumber
\end{equation}

For nonoverlapping configurations $R>2a$, the multipole matrix elements are
given by formulae (\ref{Oseen mult in R pom}), (\ref{mu coef}), and
(\ref{int three bessels nonov}). Therefore in this regime $g_{ll^{\prime}}%
^{j}\left(  R\right)  $ is proportional to $1/R^{l+l^{\prime}+1}$.

For overlapping configurations, $R<2a$, the multipole matrix elements are
given by formulae (\ref{Oseen mult in R pom}), (\ref{mu coef}), and
(\ref{int three bessels ov}) with equations (\ref{abc coef}). For small $l$
and $l^{\prime}$, analysis of the poles in Euler gamma functions and in
Pochhammer symbols (which may also be expressed by Euler gamma functions)
reveals that for overlapping configurations $g_{ll^{\prime}}^{j}\left(
R\right)  $ is a polynomial with respect to $R$. We observe that the degree of
the polynomial is $l+l^{\prime}+1$. For example%
\begin{equation}
g_{1,1}^{0}\left(  R<2a\right)  =-\frac{\left(  -2a+R\right)  ^{2}\left(
4a+R\right)  }{16\sqrt{3}},
\end{equation}%
\begin{equation}
g_{2,3}^{3}\left(  R<2a\right)  =-7\frac{\left(  -4a^{2}R+R^{3}\right)  ^{2}%
}{256\sqrt{3}}.
\end{equation}

\section{Summary}

In this article we calculate the multipole matrix elements of Green function
for Laplace equation. The elements are defined by equation (\ref{def G multi}%
). The expression for nonoverlapping configurations, i.e. for $R>2a$, is known in
the literature and can be inferred e.g. from the reference
\cite{sheu1990simulation}. The new contribution is the calculation of
expression (\ref{def G multi}) for overlapping configurations, i.e. for
$R<2a$. In this case one can find related considerations for the lowest
multipole numbers \cite{felderhof1983clausius}.

It is worth mentioning that the method which we use to calculate multipole
matrix for Laplace equation can be generalized for other cases, because the
considerations rely on simplification of the Fourier transform. For example,
the method can be generalized to the multipole matrix elements of
Green function for Stokes equations in hydrodynamics
\cite{marysiaElekFragment, cichocki2000friction, felderhof1989displacement}.
In this case overlapping configurations of multipole Green function for the
lowest multipoles have been considered recently in the literature
\cite{wajnryb2013generalization}.

\begin{acknowledgments}
K.M. has been supported by MNiSW grant IP2012 041572, and, at the earlier
stage of the research, also acknowledged support by the Foundation for Polish
Science (FNP) through the TEAM/2010-6/2 project, co-financed by the EU
European Regional Development Fund.
\end{acknowledgments}

\bibliographystyle{aipnum4-1}
\bibliography{baza_artukolow,ksiazki}

\end{document}